\documentclass[allclo]{FBSart}
\usepackage{amsfonts}
\usepackage{amssymb}
\usepackage{epsfig}

\title{Electromagnetic form factor via Minkowski and Euclidean
Bethe-Salpeter amplitudes}
\author{V.A. Karmanov$^{a}$,  J. Carbonell$^{b}$ and M.
Mangin-Brinet$^{b}$}
 \institute{$^a$Lebedev Physical Institute, Leninsky
Prospekt 53, 119991 Moscow, Russia\\ $^b$LPSC, 53 Avenue des
Martyrs, 38026 Grenoble, France}

\runningauthor{V.A. Karmanov,  J. Carbonell and M. Mangin-Brinet}
\runningtitle{Electromagnetic form factor via Minkowski and
Euclidean Bethe-Salpeter amplitudes}
\begin{document}

\maketitle
\begin{abstract}
The electromagnetic form factors (FF's) calculated through
Euclidean Bethe-Salpeter (BS) amplitude and through the
light-front (LF) wave function are compared with the one found
using the BS amplitude in Minkowski space. The FF expressed
through the Euclidean BS amplitude (both within and without static
approximation) considerably differs from the Minkowski one,
whereas FF found in the LF approach is almost indistinguishable
from it.
\end{abstract}

The electromagnetic FF's of a composite system are naturally
expressed through the BS amplitude \cite{bs} in Minkowski space.
However, finding the Minkowski BS amplitude by a direct numerical
resolution of the BS equation is complicated by its singularities.
For this reason, the BS equation is usually solved in the
Euclidean space, where singularities are absent. The total mass of
the system found this way is the same than in Minkowski space, but
the corresponding Euclidean BS amplitude drastically differs from
the Minkowski one. In terms of the Euclidean BS amplitude, the
FF's have been so far obtained only in the framework of the so
called static approximation \cite{zt}. The Euclidean BS amplitude
is first calculated in the rest frame. However, after scattering
the system obtains a recoil momentum. In the static approximation,
the BS amplitude at non zero momentum is found from the rest frame
one by a non-relativistic boost. Calculations beyond the static
approximation require an analytical continuation of the solution
in the complex plane, which is a very unstable procedure. Besides,
in full complex plane (not only on the imaginary axis) the BS
amplitude is still singular. In addition, the Wick rotation of the
integration contour, needed to express FF's through the Euclidean
BS amplitude, is not justified.

In the recent papers \cite{kc06}, a new method for solving the
Bethe-Salpeter equation was found. It is based on the Nakanishi
representation \cite{nakanishi} of the BS amplitude:
\begin{equation}\label{eq1}
\Phi(k,p)=\int_{-1}^1dz\int_0^{\infty}d\gamma
\frac{-ig(\gamma,z)}{\left[\gamma+m^2 -\frac{1}{4}M^2-k^2-p\cdot
k\; z-i\epsilon\right]^3}\ .
\end{equation}
An integral equation for the non-singular function $g(\gamma,z)$
is derived and solved numerically. Substituting this solution in
(\ref{eq1}), we can find the BS amplitude both in Minkowski and in
Euclidean spaces (taking simply real or imaginary values of
$k_0$), as well as the LF wave function:
\begin{equation}\label{eq2}
\psi(k_{\perp},x)
=\frac{1}{\sqrt{4\pi}}\int_0^{\infty}\frac{x(1-x)g(\gamma,1-2x)d\gamma}
{\Bigl(\gamma+k_{\perp}^2+m^2-x(1-x)M^2\Bigr)^2}\ .
\end{equation}
The method is valid for any kernel given by irreducible Feynman
graphs. These solutions allow us to calculate and compare the
electromagnetic FF's of a spinless particle system using different
approaches. All the calculations given below have been done with
the BS amplitude found for the ladder+cross ladder kernel. The
constituent mass $m=1$, the exchange mass $\mu=0.5$ and the
coupling constant provides the binding energy $B=1$.
\vspace{0.7cm}
\begin{figure}[h!]
\begin{center}
\includegraphics[width=8cm]{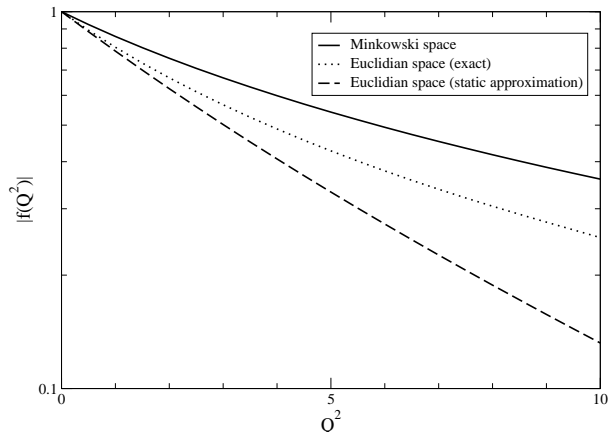}
\vspace{-0.5cm}
\caption{Solid curve: exact FF (BS in Minkowski).  Dotted curve:
FF with "exact" (complex) Euclidean BS amplitude. Dashed curve: FF
in static approximation. \label{fig1}}
\end{center}
\end{figure}

In fig. \ref{fig1} we present the FF's for $Q^2 \le 10\; m^2$
("JLab domain"). The solid curve shows the FF found through the
Minkowski space amplitude (\ref{eq1}). The dotted curve is the FF
calculated  with "exact" (complex) Euclidean BS amplitude, without
static approximation. It is obtained from the corresponding
Minkowski space one by the replacement of the integration variable
$k_0=ik_4$ by real $k_4$. This replacement corresponds to the
counter-clock-wise rotation of the integration contour from the
position $-\infty < k_0 < \infty$ to the position $-i\infty < k_0
< i\infty$, neglecting the singularities in the complex plane (if
any) which could be crossed by the contour. The difference between
curves shows that indeed some singularities are missed and,
therefore, the Wick rotation in the variable $k_0$ results in an
inaccuracy. The consequence of the static approximation (dashed
curve) is an additional error, increasing with $Q^2$.
\begin{figure}[htbp]
\vspace{-0mm}
\begin{center}
\includegraphics[width=8cm]{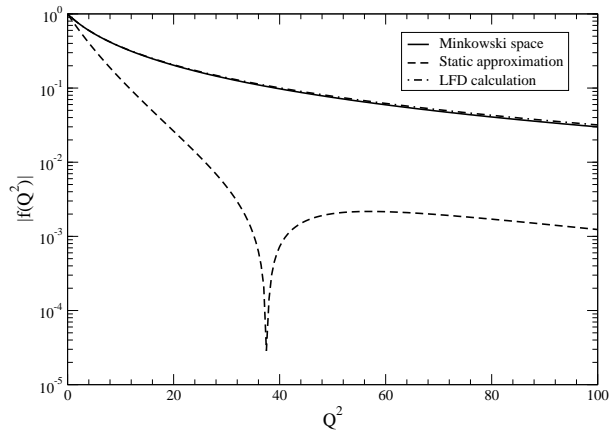}
\end{center}
\vspace{-0.7cm} \caption{Solid curve: exact FF (BS in Minkowski).
Dashed curve: FF in static approximation. Dot-dashed: FF
calculated through the LF wave function.} \label{fig2}
\end{figure}

Fig. \ref{fig2} shows the comparison \cite{KCM07} between the FF
calculated in Minkowski space (solid), in the static approximation
(dashed) and the one calculated through the LF wave function
(\ref{eq2}) (dot-dashed), for high momentum transfer. In this
domain the static FF is smaller than the Minkowski one by a factor
10 or more. A zero appears in the static FF at $Q^2\approx 38 \;
m^2$, which is an artefact of the approximation, since it is
absent in the exact (Minkowski space) FF. Regarding the LF FF, it
well coincides with the Minkowski space calculation.

We conclude that the FF calculated through the Euclidean space BS
amplitude is largely approximate, because of an error resulting
from the unjustified Wick rotation of the integration contour.
Static approximation considerably increases the error. On the
contrary, FF calculated through the LF wave function is almost
indistinguishable from the one found through the Minkowski space
BS amplitude.

Note that in contrast to the Minkowski space BS amplitude, the LF
wave function is not singular and it can be found directly, from
the corresponding 3D LF equation \cite{cdkm}, without using any BS
formalism and eq. (\ref{eq2}). This  advantage, together with very
exact result for FF, is one of the reasons which make the application
of the LF approach to the electromagnetic FF's rather attractive.


\end{document}